\documentclass[prl,showpacs,superscriptaddress,reprint]{revtex4-1}
\usepackage[T1]{fontenc}
\usepackage{amsmath}
\usepackage{amssymb}
\usepackage{graphicx}
\usepackage[caption=false]{subfig}

\makeatletter

\begin{document}
\global\long\def\ha{\hat{A}}
 \global\long\def\haa{\hat{a}}
 \global\long\def\hp{\hat{\sigma}_{z}}
 \global\long\def\hf{\hat{F}}
 \global\long\def\hr{\hat{R}}
 \global\long\def\re{\mathrm{Re}}
 \global\long\def\et{\mathrm{e}^{\mathrm{i}\theta}}
 \global\long\def\at{\ha_{\mathrm{total}}}
 \global\long\def\im{\mathrm{Im}}
 \global\long\def\d{\mathrm{d}}
 \global\long\def\net{\mathrm{e}^{-\mathrm{i}\theta}}
 \global\long\def\i{\mathrm{i}}
\global\long\def\e{\mathrm{e}}

\title{Entanglement-assisted weak value amplification}

\author{Shengshi Pang}

\affiliation{Department of Electrical Engineering, University of Southern California,
Los Angeles, CA 90089, USA.}

\author{Justin Dressel}

\affiliation{Department of Electrical Engineering, University of California, Riverside,
CA 92521, USA.}

\author{Todd A. Brun}

\affiliation{Department of Electrical Engineering, University of Southern California,
Los Angeles, CA 90089, USA.}
\begin{abstract}
Large weak values have been used to amplify the sensitivity of a linear
response signal for detecting changes in a small parameter, which
has also enabled a simple method for precise parameter estimation.
However, producing a large weak value requires a low postselection
probability for an ancilla degree of freedom, which limits the utility
of the technique. We propose an improvement to this method that uses
entanglement to increase the efficiency. We show that by entangling
and postselecting $n$ ancillas, the postselection probability can
be increased by a factor of $n$ while keeping the weak value fixed
(compared to $n$ uncorrelated attempts with one ancilla), which is
the optimal scaling with $n$ that is expected from quantum metrology.
Furthermore, we show the surprising result that the quantum Fisher
information about the detected parameter can be almost entirely preserved
in the postselected state, which allows the sensitive estimation to
approximately saturate the optimal quantum Cram\'{e}r-Rao bound. To
illustrate this protocol we provide simple quantum circuits that can
be implemented using current experimental realizations of three entangled
qubits. 
\end{abstract}


\maketitle
Weak value amplification is an enhanced detection scheme that was
first suggested by Aharonov, Albert, and Vaidman \cite{AAV}. (See
\cite{Kofman2012} and \cite{justin} for recent reviews.) The scheme
exploits the fact that postselecting the weak measurement of an ancilla
can produce a linear detector response with an anomalously high sensitivity
to small changes in an interaction parameter. The sensitivity arises
from coherent ``super-oscillatory'' interference in the ancilla
\cite{Berry2012}, which is controlled by the choice of preparation
and postselection of the ancilla. The price that one pays for this
increase in sensitivity is a reduction in the potential signal (and
thus the potential precision of any estimation) due to the postselection
process \cite{Zhu2011,Tanaka2013,Ferrie2013,Knee2013,Combes2013}.
Nevertheless, by using this technique one can still consistently recover
a large fraction of the maximum obtainable signal in a relatively
simple way \cite{Starling2009,Feizpour2011}. The relevant information
is effectively concentrated into the small set of rarely postselected
events \cite{Jordan2013}.

A growing number of experiments have successfully used weak value
amplification to precisely estimate a diverse set of small physical
parameters, including beam deflection (to picoradian resolution) \cite{science spin hall,signaltonoise2,Turner2011,Pfeifer2011,Hogan2011,Zhou2012,Zhou2013,Jayaswal2014},
frequency shifts \cite{precision}, phase shifts \cite{Starling2010,Xu2013},
angular shifts \cite{Magana2013}, temporal shifts \cite{Strubi2013,Viza2013}, and temperature shifts
\cite{Egan2012}. More experimental schemes have also been proposed
\cite{proposal-chargesensing,proposal-electron spin,proposal-phaseshift,proposal-Tomography of Many-Body Weak Value,proposal-wu-marek,proposal-fermion,Susa2012,Hayat2014}.
These experimental results have shown remarkable resiliance to the
addition of temporally-correlated noise, such as beam jitter \cite{Jordan2013}.
Moreover, some of these experiments have reported precision near the
standard quantum limit, which is surprising due to the intrinsic postselection
loss. These observations have prompted the question of whether the
amplification technique can be improved further by combining it with
other metrology techniques. One such improvement that has been proposed
is to recycle the events that were discarded by the postselection
back into the measurement \cite{Dressel2013}. Another investigation
has shown the that in certain cases it may even be possible to achieve
precision near the optimal Heisenberg limit using seemingly classical
resources \cite{Zhang2013}.

In this Letter we supplement these efforts by asking whether adding
quantum resources to the weak value amplification procedure can also
improve the efficiency of the technique. We find that using entangled
ancilla preparations and postselections does indeed provide such an
improvement. That is, the postselection probability can be increased
while preserving the amplification factor, which effectively decreases
the number of discarded events required to achieve the same sensitivity.
Alternatively, one can enhance the amplification directly while preserving
the same postselection probability. We show that these improvements
scale optimally as the number of entangled ancillas increases; however,
using even a small number of entangled ancillas provides a notable
improvement. Moreover, we show that nearly all the quantum Fisher
information about the estimated parameter can be preserved in the
rarely postselected state, which allows the parameter estimation to
nearly saturate the quantum Cram\'{e}r-Rao bound in the weak value
regime.

As a concrete proposal that demonstrates this optimal scaling, we
consider using $n$ entangled ancilla qubits \cite{brun} to estimate
a small controlled phase applied to a meter qubit. Recent experiments
with optical \cite{experiment-weak value}, solid-state \cite{Groen2013,Campagne2013}
and NMR \cite{Lu2013} systems have already verified the weak value
effect using one or two qubits. As such, we provide a simple set of
similar quantum circuits that can be implemented experimentally in
a straightforward way using only three physical qubits.

\emph{Weak value amplification}.--- As a brief review, recall that
for a typical weak value amplification experiment one uses an interaction
Hamiltonian of the form 
\begin{equation}
\hat{H}_{\mathrm{int}}=\hbar g\ha\otimes\hf\,\delta(t-t_{0}),\label{eq:hamiltonian}
\end{equation}
where $\ha$ is an ancilla observable, $\hf$ is a meter observable,
and $g$ is the small coupling parameter that one would like to estimate.
The time factor $\delta(t-t_{0})$ indicates that the interaction
between the ancilla and the meter is impulsive, i.e., happening on
a much faster timescale than the natural evolution of both the ancilla
and the meter. Importantly for our discussion, we leave the dimension
of $\ha$ arbitrary.

An experimenter prepares the meter in a pure state $|\phi\rangle$
and the ancilla in a pure initial state $|\Psi_{i}\rangle$, then
weakly couples them using the interaction Hamiltonian of Eq.~\eqref{eq:hamiltonian},
and then postselects the ancilla into a pure final state $|\Psi_{f}\rangle$,
discarding the events where the postselection fails. This procedure
effectively prepares an \emph{enhanced} meter state that includes
the effect of the ancilla $|\phi'\rangle=\hat{M}|\phi\rangle/||\hat{M}|\phi\rangle||$,
which we write here in terms of a Kraus operator $\hat{M}=\langle\Psi_{f}|\exp(-\i g\ha\otimes\hf)|\Psi_{i}\rangle$.
Averaging a meter observable $\hr$ using this updated meter state
yields $\langle\hr\rangle_{|\phi'\rangle}=\langle\phi|\hat{M}^{\dagger}\hr\hat{M}|\phi\rangle/\langle\phi|\hat{M}^{\dagger}\hat{M}|\phi\rangle$.

For small $g$, this observable average is well approximated by the
following second-order expansion \cite{DiLorenzo2012,Kofman2012}
\begin{equation}
\langle\hr\rangle_{|\phi'\rangle}\approx\frac{2g\,\text{Im}(\alpha\, A_{w})+g^{2}\beta|A_{w}|^{2}}{1+g^{2}\sigma^{2}|A_{w}|^{2}},\label{eq:secondorder}
\end{equation}
where $\alpha=\langle\hr\hf\rangle_{|\phi\rangle}$, $\beta=\langle\hf\hr\hf\rangle_{|\phi\rangle}$,
and $\sigma^{2}=\langle\hf^{2}\rangle_{|\phi\rangle}$ are correlation
parameters that are fixed by the choice of meter observables and initial
state, while 
\begin{equation}
A_{w}=\frac{\langle\Psi_{f}|\ha|\Psi_{i}\rangle}{\langle\Psi_{f}|\Psi_{i}\rangle}\label{eq:1}
\end{equation}
is a complex \emph{weak value} controlled by the ancilla \cite{AAV}.
Note that we have assumed that the initial meter state is unbiased
$\langle\hf\rangle_{|\phi\rangle}=\langle\hr\rangle_{|\phi\rangle}=0$
to obtain the best response.

Most amplification experiments operate in the linear response regime
where the second-order terms in Eq.~\eqref{eq:secondorder} can be
neglected, which produces \cite{josza} 
\begin{equation}
\langle\hr\rangle_{|\phi'\rangle}\approx2g\left[\text{Re}A_{w}\,\text{Im}\alpha+\text{Im}A_{w}\,\text{Re}\alpha\right].\label{eq:linear}
\end{equation}
This linear relation shows how a large ancilla weak value can amplify
the sensitivity of the meter for detecting small changes in $g$.

For concreteness, we consider a reference case when the meter is a
qubit. State-of-the-art quantum computing technologies can already
realize single qubit unitary gates and two qubit \texttt{CNOT} and
controlled rotation gates with high fidelity (e.g., \cite{experiment-weak value,Groen2013,Campagne2013,Lu2013,Reed2012,Chow2012,Murch2013,Zhong2013}),
so this example can be readily tested in the laboratory. The meter
qubit is prepared in the state $|\phi\rangle=|+\rangle=(|0\rangle+|1\rangle)/\sqrt{2}$.
The Pauli $Z$-operator $\hat{\sigma}_{z}=\hf=\hr$ will serve as
both meter observables. These choices fix the constants $\alpha=1$,
$\beta=0$, and $\sigma^{2}=1$ in Eq.~\eqref{eq:secondorder}, yielding
the meter response 
\begin{equation}
\langle\hat{\sigma}_{z}\rangle_{|+'\rangle}\approx\frac{2g\,\text{Im}A_{w}}{1+g^{2}\,|A_{w}|^{2}}.\label{eq:qubitresponse}
\end{equation}
The nonlinearity in the denominator regularizes the detector response,
placing a strict upper bound of $g|A_{w}|<1$ on the magnitudes that
are useful for amplification purposes. The meter has a linear response
in a more restricted range of roughly $g|A_{w}|<1/10$. In practice,
one typically assumes that $g|A_{w}|\ll1$.

\begin{figure}
\includegraphics[width=0.8\columnwidth]{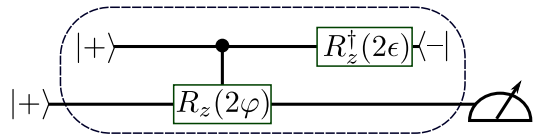} \caption{Quantum circuit that simulates the weak value amplification of a small
parameter $\varphi$. A meter qubit is prepared in the state $|+\rangle=R_{y}(\pi/2)|0\rangle=(|0\rangle+|1\rangle)/\sqrt{2}$.
An ancilla qubit is prepared in the same state $|\Psi_{i}\rangle=|+\rangle$.
The ancilla is used as a control for a $Z$-rotation $R_{z}(2\varphi)$
of the meter, which simulates the unitary $\hat{U}=\exp(-\i\varphi\ha\otimes\hat{\sigma}_{z}/2)$
with $\ha=\hat{\sigma}_{z}$. The ancilla is then postselected in
the nearly orthogonal state $\langle\Psi_{f}|=\langle-|R_{z}^{\dagger}(2\epsilon)=\langle0|R_{y}^{\dagger}(-\pi/2)R_{z}^{\dagger}(2\epsilon)=(\langle0|\e^{\i\epsilon}-\langle1|\e^{-\i\epsilon})/\sqrt{2}$
with probability $P_{s}\approx\epsilon^{2}$ by performing two rotations,
measuring in the $Z$-basis, and keeping only the $\langle0|$ events.
Finally, the meter qubit is measured in the $Z$-basis, which yields
the linear response $\langle\hat{\sigma}_{z}\rangle_{+'}\approx\varphi\,\text{Im}A_{w}$
that is amplified by the weak value $A_{w}\approx\i/\epsilon$. The
probability for a single success of this circuit after $n$ attempts,
$P_{s}^{(n)}=1-(1-P_{s})^{n}\approx n\,\epsilon^{2}$, is approximately
linear in $n$.}

\label{fig:circuit} 
\end{figure}

As detailed in Figure~\ref{fig:circuit}, we couple a single ancilla
qubit to the meter using a controlled-$Z$ rotation by a small angle
$2\varphi$, which sets $g=\varphi/2$ and $\ha=\hat{\sigma}_{z}$.
The ancilla is initialized in the state $|\Psi_{i}\rangle=|+\rangle$
and postselected in the nearly orthogonal state $|\Psi_{f}\rangle=R_{z}(2\epsilon)|-\rangle=(\e^{-\i\epsilon}|0\rangle-\e^{\i\epsilon}|1\rangle)/\sqrt{2}$
with a probability $P_{s}=\sin^{2}(\epsilon)\approx\epsilon^{2}$,
which produces the weak value $A_{w}=\i\cot(\epsilon)\approx\i/\epsilon$.
The offset angle $\epsilon$ of the postselection must satisfy $\varphi/2<\epsilon<\pi/4$
for amplification, and $5\varphi<\epsilon<\pi/4$ for linear response.

\emph{Postselection probability}.--- While the weak value has the
marvelous ability to effectively amplify the small parameter $g$
in a simple way, it also has a shortcoming of low efficiency. That
is, for a large weak value $A_{w}$, Eq.~(\ref{eq:1}) indicates
that $\langle\Psi_{f}|\Psi_{i}\rangle$ must be small. This implies
that the ancilla postselection probability is also small, since it
approximates 
\begin{equation}
P_{s}\approx|\langle\Psi_{f}|\Psi_{i}\rangle|^{2}\label{eq:5}
\end{equation}
for small $g$. Therefore, the larger $A_{w}$ is, the less likely
it is that one can successfully postselect the ancilla and prepare
the amplified meter state $|\phi'\rangle$.

We now show that adding quantum resources to the ancilla can improve
this efficiency while keeping the amplification factor of the weak
value $A_{w}$ the same. Specifically, we consider coupling $n$ entangled
ancillas to the meter simultaneously. To make a fair comparison with
the uncorrelated case, the probability of successfully postselecting
$n$ entangled ancillas once should show an improvement over the probability
of successfully postselecting a single ancilla once after $n$ independent
attempts. The latter probability has linear scaling in $n$ when $P_{s}$
is small 
\begin{equation}
P_{s}^{(n)}=1-(1-P_{s})^{n}\approx nP_{s}.\label{eq:ref}
\end{equation}
We will see that entangled ancillas can achieve \emph{quadratic} scaling
with $n$, which improves the postselection efficiency by a factor
of $n$.

\begin{figure}
\includegraphics[width=0.8\columnwidth]{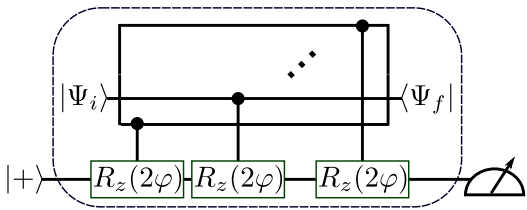} \caption{Quantum circuit that simulates the entanglement-assisted weak value
amplification of a small parameter $\varphi$. As in Figure~\ref{fig:circuit},
a meter qubit is prepared in the state $|+\rangle$, while $n$ ancilla
qubits are prepared in a entangled state $|\Psi_{i}\rangle$. Each
ancilla is then used as a control for a $Z$-rotation $R_{z}(2\varphi)$
of the meter, simulating the unitary $\hat{U}=\exp(-\i\varphi\ha\otimes\hat{\sigma}_{z}/2)$
with $\ha$ being the sum of ancilla observables $\hat{\sigma}_{z}$.
The ancillas are then postselected in an entangled state $|\Psi_{f}\rangle$,
and the meter qubit is measured in the $Z$-basis, yielding the linear
response $\langle\hat{\sigma}_{z}\rangle_{+'}\approx\varphi\,\text{Im}A_{w}$
amplified by a joint weak value $A_{w}$.}

\label{fig:circuitentangled} 
\end{figure}

To show this improvement, we couple the meter to $n$ identical single-ancilla
observables $\haa$ using the interaction in Eq.~\eqref{eq:hamiltonian},
which effectively couples the meter to a single joint ancilla observable
\begin{equation}
\ha=\ha_{1}+\cdots+\ha_{n},\label{eq:ancillatot}
\end{equation}
where $\ha_{k}=\hat{1}\otimes\cdots\haa\cdots\otimes\hat{1}$ is shorthand
for the observable $\haa$ of the $k$\textsuperscript{th} ancilla.
Notably the minimum and maximum eigenvalues of this joint observable,
$\Lambda_{\min(\max)}=n\lambda_{\min(\max)}$, are determined by the
eigenvalues of $\haa$. Similarly, the corresponding eigenstates are
product states of the eigenstates of $\haa$: $|\Lambda_{\min(\max)}\rangle=|\lambda_{\min(\max)}\rangle^{\otimes n}$.
The $n$ ancillas will be collectively prepared in a joint state $|\Psi_{i}\rangle$
and then postselected in a joint state $|\Psi_{f}\rangle$ to produce
a joint weak value amplification factor $A_{w}$, just as in Eq.~\eqref{eq:1}.
An example circuit that implements this procedure with qubits is illustrated
in Figure~\ref{fig:circuitentangled}.

The ability to improve the postselection efficiency hinges upon the
fact that there can be different choices of $|\Psi_{i}\rangle$ and
$|\Psi_{f}\rangle$ that will produce the same weak value $A_{w}$.
However, these different choices will generally produce different
postselection probabilities. Therefore, among these different choices
of joint preparations and postselections there exists an optimal choice
that maximizes the postselection probability.

We find this optimum in two steps. First, we maximize the postselection
probability over all possible postselections $|\Psi_{f}\rangle$ while
keeping the weak value $A_{w}$ and the preparation $|\Psi_{i}\rangle$
fixed. Second, we maximize this result over all preparations $|\Psi_{i}\rangle$.

To perform the first maximization, note that Eq.~(\ref{eq:1}) implies
$\langle\Psi_{f}|(\ha-A_{w})|\Psi_{i}\rangle=0$, so $|\Psi_{f}\rangle$
must be orthogonal to $(\ha-A_{w})|\Psi_{i}\rangle$. This gives a
constraint on the possible postselections $|\Psi_{f}\rangle$, so
the maximization of $P_{S}$ in Eq.~(\ref{eq:5}) should be taken
over the subspace $\mathcal{V}^{\perp}$ orthogonal to $(\ha-A_{w})|\Psi_{i}\rangle$.
As shown in the Supplementary Material \cite{supplement}, the result
of this maximization approximates 
\begin{equation}
\max_{|\Psi_{f}\rangle\in\mathcal{V}^{\perp}}P_{s}\approx\frac{\text{Var}(\ha)_{|\Psi_{i}\rangle}}{|A_{w}|^{2}},\label{eq:6}
\end{equation}
where $\text{Var}(\ha)_{|\Psi_{i}\rangle}=\langle\Psi_{i}|\ha^{2}|\Psi_{i}\rangle-[\langle\Psi_{i}|\ha|\Psi_{i}\rangle]^{2}$
is the variance of $\ha$ in the initial state. This approximation
applies when the weak value is larger than any eigenvalue of $\ha$:
$|\Lambda|\ll|A_{w}|<1/g$. However, since $\Lambda_{\min(\max)}=n\lambda_{\min(\max)}$,
we must be careful to fix $|A_{w}|$ to be at least $n$ times larger
than the eigenvalues of $\haa$.

Now we consider maximizing the variance over an arbitrary initial
state $|\Psi_{i}\rangle$, which produces \cite{quantum metrology 2}
\begin{align}
\max_{|\Psi_{i}\rangle}\text{Var}(\ha)_{|\Psi_{i}\rangle} & =\frac{n^{2}}{4}(\lambda_{\max}-\lambda_{\min})^{2},\label{eq:8}
\end{align}
showing \emph{quadratic} scaling with $n$. Therefore, according to
Eq.~\eqref{eq:6} the maximum postselection probability also scales
quadratically with $n$, showing a factor of $n$ improvement over
the linear scaling of the uncorrelated ancilla attempts in Eq.~\eqref{eq:ref}.

\begin{figure}
\begin{centering}
\includegraphics[width=0.5\columnwidth]{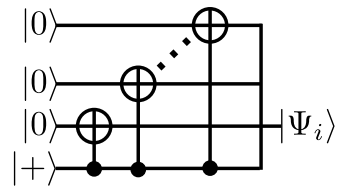} 
 
\par\end{centering}

\caption{Quantum circuit to prepare the optimal entangled preparation for $n$
ancilla qubits. The state $|\Psi_{i}\rangle=(|0\rangle^{\otimes n}+|1\rangle^{\otimes n})/\sqrt{2}$
is prepared from a single $|+\rangle$ state by a sequence of \texttt{CNOT}
gates. Due to this construction, we note that the ordering of the
two-qubit gates in Figs.~\ref{fig:circuitentangled}, \ref{fig:preparation},
and \ref{fig:postselection} can be further optimized to pre- and
postselect $(n-1)$ of the ancilla qubits sequentially, which allows
the $n$-qubit entangled ancilla to be practically simulated using
only three physical qubits. }

\label{fig:preparation} 
\end{figure}

The preparation states that show this quadratic scaling of the variance
have the maximally entangled form \cite{quantum metrology 2} 
\begin{align}
|\Psi_{i}\rangle & =\frac{1}{\sqrt{2}}(|\lambda_{\max}\rangle^{\otimes n}+\et|\lambda_{\min}\rangle^{\otimes n}),\label{eq:9}
\end{align}
where $\et$ is an arbitrary relative phase. We provide a simple circuit
to prepare such a state for $n$ qubits in Figure~\ref{fig:preparation},
choosing $\theta=0$.

\begin{figure}
\begin{centering}
\subfloat[Postselection maximizing $P_{s}$]{ \includegraphics[width=0.6\columnwidth]{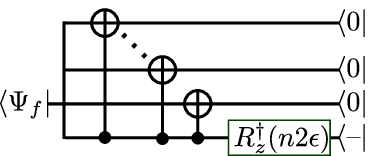}

}
\par\end{centering}

\begin{centering}
\subfloat[Postselection maximizing $A_{w}$]{ \includegraphics[width=0.6\columnwidth]{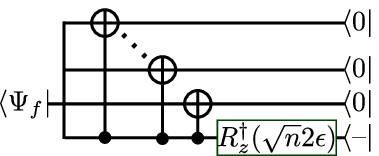}

}
\par\end{centering}

\caption{Quantum circuits for attaining optimal postselections, using the preparation
in Figure~\ref{fig:preparation}. (a) Keeping $A_{w}\approx\i/\epsilon$
fixed and maximizing $P_{s}$ produces the entangled postselection
$\langle\Psi_{f}|=\langle0|^{\otimes n}\e^{\i n\epsilon}-\langle1|^{\otimes n}\e^{-\i n\epsilon}$
with $P_{s}\approx n^{2}\epsilon^{2}$, which is a factor of $n$
larger than the single ancilla $P_{s}^{(n)}$ in Figure~\ref{fig:circuit}.
This postselection can be implemented as a sequence of \texttt{CNOT}
gates and a rotation of the last qubit by $R_{z}^{\dagger}(n2\epsilon)$
and $R_{y}^{\dagger}(-\pi/2)$ before measuring all qubits in the
$Z$-basis and keeping only $\langle0|$ events. For small $\epsilon$
this state is equivalent to Eq.~\eqref{eq:postselect}. (b) Keeping
$P_{s}=P_{s}^{(n)}\approx n\epsilon^{2}$ and maximizing $A_{w}$
produces a similar state $\langle\Psi_{f}|=\langle0|^{\otimes n}\e^{\i\sqrt{n}\epsilon}-\langle1|^{\otimes n}\e^{-\i\sqrt{n}\epsilon}$
with $A_{w}\approx\i\sqrt{n}/\epsilon$, which is a factor of $\sqrt{n}$
larger than $A_{w}$ in Figure~\ref{fig:circuit}. }

\label{fig:postselection} 
\end{figure}

According to the derivation in the Supplementary Material \cite{supplement},
the corresponding postselection states that maximize the postselection
probability are 
\begin{align}
|\Psi_{f}\rangle & \propto-(n\lambda_{\min}-A_{w}^{*})|\lambda_{\max}\rangle^{\otimes n}\label{eq:postselect}\\
 & \qquad+\et(n\lambda_{\max}-A_{w}^{*})|\lambda_{\min}\rangle^{\otimes n},\nonumber 
\end{align}
which explicitly depend on the chosen value of $A_{w}$. We also provide
a simple circuit to implement this postselection with $n$ qubits
in Figure~\ref{fig:postselection}(a).

\emph{Weak value scaling}.--- So far we have shown that we can increase
the postselection probability by a factor of $n$ when the weak value
is kept fixed. Alternatively, we can hold the postselection probability
fixed to increase the maximum weak value by factor of $\sqrt{n}$.

Given a specific postselection probability $P_{s}$, the postselected
state $|\Psi_{f}\rangle$ must have the form 
\begin{equation}
|\Psi_{f}\rangle=\sqrt{P_{s}}|\Psi_{i}\rangle+\sqrt{1-P_{s}}\et|\Psi_{i}^{\perp}\rangle,
\end{equation}
where $|\Psi_{i}^{\perp}\rangle$ is an arbitrary state orthogonal
to $|\Psi_{i}\rangle$. This implies that we can write the weak value
in Eq.~\eqref{eq:1} as 
\begin{equation}
A_{w}=\langle\Psi_{i}|\ha|\Psi_{i}\rangle+\sqrt{\frac{1-P_{s}}{P_{s}}}\net\langle\Psi_{i}^{\perp}|\ha|\Psi_{i}\rangle.
\end{equation}
For large $A_{w}$ and small $P_{s}$, then we can approximately neglect
the first term. Since $\et$ is arbitrary, we can also assume that
$\langle\Psi_{i}^{\perp}|\ha|\Psi_{i}\rangle$ is positive. The maximum
$\langle\Psi_{i}^{\perp}|\ha|\Psi_{i}\rangle$ can be achieved when
$|\Psi_{i}^{\perp}\rangle$ is parallel to the component of $\ha|\Psi_{i}\rangle$
in the complementary subspace orthogonal to $|\Psi_{i}\rangle$. This
choice produces $\langle\Psi_{i}^{\perp}|\ha|\Psi_{i}\rangle=\|\ha|\Psi_{i}\rangle-|\Psi_{i}\rangle\langle\Psi_{i}|\ha|\Psi_{i}\rangle\|=[\text{Var}(\ha)_{|\Psi_{i}\rangle}]^{1/2}$.
Therefore, the largest weak value that can be obtained from the initial
state $|\Psi_{i}\rangle$ with a small postselection probability $P_{s}$
will approximate 
\begin{equation}
\max|A_{w}|\approx\sqrt{\frac{\text{Var}(\ha)_{|\Psi_{i}\rangle}}{P_{s}}}.\label{eq:1-1}
\end{equation}

That is, the variance controls the scaling for the maxima of both
$P_{s}$ and $A_{w}$. Comparing Eqs.~\eqref{eq:6} and \eqref{eq:1-1},
it follows that if $P_{s}$ can be improved by a factor of $n$, then
it is also possible to improve $A_{w}$ by a factor of $\sqrt{n}$.
Furthermore, maximizing the variance produces the same initial state
as Eq.~\eqref{eq:9}, so the only difference between maximizing $P_{s}$
and $A_{w}$ is the choice of postselection state. We provide a simple
circuit to implement this alternative postselection with $n$ qubits
in Figure~\ref{fig:postselection}(b).

\emph{Fisher information}.--- An improvement factor of $\sqrt{n}$
in the estimation precision is the best that we can expect from using
entangled ancillas, according to well-known results from quantum metrology
\cite{quantum metrology 2,quantum metrology 3,footnote}. We are thus
faced with the conundrum of how such a rare postselection can possibly
show such optimal scaling with $n$. After all, most of the (potentially
informative) data is being discarded by the postselection.

To understand this behavior, we compare the quantum Fisher information
$I(g)$ about $g$ contained in the post-interaction state $|\Phi_{g}\rangle=\exp(-\i g\hat{A}\otimes\hat{F})|\Psi_{i}\rangle|\phi\rangle$
to the Fisher information $I'(g)$ that remains in the postselected
state $\sqrt{P_{s}}|\phi'\rangle$. As detailed in the Supplementary
Material \cite{supplement}, in the linear response regime $g|A_{w}|\text{Var}(\hf)^{\frac{1}{2}}\ll1$
with an initially unbiased meter $\langle\hat{F}\rangle_{|\phi\rangle}=0$,
and assuming a fixed $P_{s}\ll1$ with maximal $|A_{w}|$, we obtain
\begin{equation}
I'(g)\approx\eta\, I(g)\,[1-|gA_{w}|^{2}\text{Var}(\hf)]\leq I(g),
\end{equation}
where $\eta=\text{Var}(\hat{A})_{|\Psi_{i}\rangle}/\langle\hat{A}^{2}\rangle_{|\Psi_{i}\rangle}$
is an efficiency factor.

Remarkably, $\eta$ can reach $1$ when $\langle\hat{A}\rangle_{|\Psi_{i}\rangle}=0$,
implying that nearly all the original Fisher information $I(g)$ can
be concentrated into one rarely obtained $|\phi'\rangle$, up to a
small reduction by $|gA_{w}|^{2}\text{Var}(\hf)\ll1$. The remaining
information is distributed among the discarded meter states, and could
be retrieved in principle \cite{Ferrie2013,Combes2013}. For the example
with $\hat{F}=\haa=\hat{\sigma}_{z}$, the initial state in Eq.~\eqref{eq:9}
yields $\eta=1$, $\text{Var}(\hf)=1$, and a total Fisher information
of $I(g)=4\langle\hat{A}^{2}\rangle_{|\Psi_{i}\rangle}=4n^{2}$ (see
the Supplementary Material \cite{supplement}). The Cram\'{e}r-Rao
bound is thus $[I'(g)]^{-\frac{1}{2}}=(1/2n)[1-|gA_{w}|^{2}]^{-\frac{1}{2}}$
for the precision of any unbiased estimation of $g=\varphi/2$ using
$|\phi'\rangle$, confirming the optimal scaling with $n$.

\emph{Conclusion}.--- In summary, we have considered using entanglement
to enhance the weak value amplification of a small parameter. If the
amplification factor is held fixed, then $n$ entangled ancillas can
improve the postselection probability by a factor of $n$ compared
to $n$ attempts with uncorrelated ancillas. This improvement in postselection
efficiency addresses a practical shortcoming of weak value amplification,
and achieves the optimal scaling with $n$ that can be expected from
quantum metrology. Indeed, we have shown that weak value amplification
can nearly saturate the quantum Cram\'{e}r-Rao bound, despite the
low efficiency of postselection. We have also provided simple quantum
circuits for the protocol that are readily implementable by existing
quantum computing architectures that possess three qubits.

\emph{Acknowledgments}.--- JD thanks Alexander Korotkov, Eyob Sete,
and Andrew Jordan for helpful discussions. This research was partially
supported by the ARO MURI grant W911NF-11-1-0268. SP and TB also acknowledge
the support from NSF grant CCF-0829870, while JD acknowledges support
from IARPA/ARO grant W91NF-10-1-0334.

\appendix

\begin{widetext}

\section{Derivation of the maximum post-selection probability}

To maximize $P_{s}\approx|\langle\Psi_{f}|\Psi_{i}\rangle|^{2}$ while
keeping $A_{w}$ and $|\Psi_{i}\rangle$ fixed, we note that the initial
state can be decomposed into a piece parallel to $(\ha-A_{w})|\Psi_{i}\rangle$
and an orthogonal piece in the complementary subspace $\mathcal{V}^{\perp}$:
\begin{align}
|\Psi_{i}\rangle & =\frac{(\ha-A_{w})|\Psi_{i}\rangle\langle\Psi_{i}|(\ha-A_{w}^{*})|\Psi_{i}\rangle}{\langle\Psi_{i}|(\ha-A_{w}^{*})(\ha-A_{w})|\Psi_{i}\rangle}+\left(|\Psi_{i}\rangle-\frac{(\ha-A_{w})|\Psi_{i}\rangle\langle\Psi_{i}|(\ha-A_{w}^{*})|\Psi_{i}\rangle}{\langle\Psi_{i}|(\ha-A_{w}^{*})(\ha-A_{w})|\Psi_{i}\rangle}\right).
\end{align}
Since $|\Psi_{f}\rangle$ must also be in $\mathcal{V}^{\perp}$ by
the definition of the weak value, it follows that the maximum $P_{s}$
can be achieved for the post-selection state parallel to the component
of $|\Psi_{i}\rangle$ in $\mathcal{V}^{\perp}$, i.e., 
\begin{equation}
|\Psi_{f}\rangle\propto|\Psi_{i}\rangle-\frac{(\ha-A_{w})|\Psi_{i}\rangle\langle\Psi_{i}|(\ha-A_{w}^{*})|\Psi_{i}\rangle}{\langle\Psi_{i}|(\ha-A_{w}^{*})(\ha-A_{w})|\Psi_{i}\rangle}.\label{eq:11}
\end{equation}
After some calculation, it follows that 
\begin{equation}
\max_{|\Psi_{f}\rangle\in\mathcal{V}^{\perp}}P_{s}=\frac{\text{Var}(\ha)_{|\Psi_{i}\rangle}}{\langle\Psi_{i}|\ha^{2}|\Psi_{i}\rangle-2\langle\Psi_{i}|\hat{A}|\Psi_{i}\rangle\re A_{w}+|A_{w}|^{2}},\label{eq:12}
\end{equation}
where $\text{Var}(\ha)_{|\Psi_{i}\rangle}=\langle\Psi_{i}|\ha^{2}|\Psi_{i}\rangle-[\langle\Psi_{i}|\ha|\Psi_{i}\rangle]^{2}$
is the variance of $\ha$ in the state $|\Psi_{i}\rangle$.

For the purposes of weak value amplification, we usually require $|A_{w}|$
to be larger than any eigenvalue of $\ha$, $|A_{w}|\gg|\Lambda|$.
Therefore, this maximum $P_{s}$ can be approximated as Eq. (9) in
the main text.

\section{Derivation of the optimal post-selection state}

As noted in the previous section, the optimal post-selection state
should be parallel to the component of $|\Psi_{i}\rangle$ in $\mathcal{V}^{\perp}$.
The post-selection probability is then controlled by the variance
$\text{Var}(\hat{A})_{|\Psi_{i}\rangle}$. This variance is maximized
for a maximally entangled initial state $|\Psi_{i}\rangle=\frac{1}{\sqrt{2}}(|\lambda_{\max}\rangle^{\otimes n}+\et|\lambda_{\min}\rangle^{\otimes n})$.
Hence, we can directly compute the optimal post-selected state to
be 
\begin{align}
|\Psi_{f}\rangle & \propto|\Psi_{i}\rangle-\frac{(\at-A_{w})|\Psi_{i}\rangle\langle\Psi_{i}|(\at-A_{w}^{*})|\Psi_{i}\rangle}{\langle\Psi_{i}|(\at-A_{w}^{*})(\at-A_{w})|\Psi_{i}\rangle}\\
 & =\frac{1}{\sqrt{2}}(|\lambda_{\max}\rangle^{\otimes n}+\et|\lambda_{\min}\rangle^{\otimes n})-\frac{1}{\sqrt{2}}((n\lambda_{\max}-A_{w})|\lambda_{\max}\rangle^{\otimes n}\nonumber \\
 & \qquad+\et(n\lambda_{\min}-A_{w})|\lambda_{\min}\rangle^{\otimes n})\frac{n\lambda_{\max}+n\lambda_{\min}-2A_{w}^{*}}{|n\lambda_{\max}-A_{w}|^{2}+|n\lambda_{\min}-A_{w}|^{2}}\nonumber \\
 & \propto(|n\lambda_{\min}-A_{w}|^{2}-(n\lambda_{\max}-A_{w})(n\lambda_{\min}-A_{w}^{*}))|\lambda_{\max}\rangle^{\otimes n}\nonumber \\
 & \qquad+\et(|n\lambda_{\max}-A_{w}|^{2}-(n\lambda_{\min}-A_{w})(n\lambda_{\max}-A_{w}^{*}))|\lambda_{\min}\rangle^{\otimes n})\nonumber \\
 & \propto-(n\lambda_{\min}-A_{w}^{*})|\lambda_{\max}\rangle^{\otimes n}+\et(n\lambda_{\max}-A_{w}^{*})|\lambda_{\min}\rangle^{\otimes n}.\nonumber 
\end{align}
This is Eq.~(12) in the main text.

\section{Quantum Fisher information}

It is important to determine just how well the weak value amplification
technique can estimate the small parameter $g$. There is some concern
that the post-selection process will lead to a substantial reduction
of the total obtainable information, since a large fraction of the
potentially usable data is being thrown away (e.g., \cite{Ferrie2013}).
To assuage these concerns, we compare the maximum Fisher information
about $g$ that can be obtained without post-selection to the Fisher
information that remains in the post-selected states used for weak
value amplification.

We first recall a few general results from the study of quantum Fisher
information. If one wishes to estimate a parameter $g$, then the
minimum standard deviation of any unbiased estimator for $g$ is given
by the \emph{quantum Cram\'{e}r-Rao bound}: $I(g)^{-1/2}$. The function
$I(g)$ is the \emph{quantum Fisher information} \cite{metrology2}
\begin{equation}
I(g)=4\frac{\d\langle\Phi_{g}|}{\d g}\frac{\d|\Phi_{g}\rangle}{\d g}-4\left|\frac{\d\langle\Phi_{g}|}{\d g}|\Phi_{g}\rangle\right|^{2},\label{eq:2}
\end{equation}
which is determined by a quantum state $|\Phi_{g}\rangle$ that contains
the information about $g$. If this state is prepared with some interaction
Hamiltonian $|\Phi_{g}\rangle=\exp(-ig\hat{H})|\Phi\rangle$ then
the Fisher information reduces to a simpler form \cite{quantum metrology 2}
\begin{equation}
I(g)=4\text{Var}(\hat{H})_{|\Phi\rangle},\label{eq:1-1}
\end{equation}
and is entirely determined by the variance of the Hamiltonian in the
pre-interaction state $|\Phi\rangle$.

\subsection{General Discussion}

In the main text, the relevant Hamiltonian with a meter observable
$\hat{F}$ is $\hat{H}=\hbar g\hat{A}\otimes\hat{F}\delta(t-t_{0})$,
where $\hat{A}$ is a sum of $n$ ancilla observables $\hat{a}$ of
dimension $d$. The joint state $|\Phi\rangle$ is also always prepared
in a product state $|\Phi\rangle=|\Psi_{i}\rangle\otimes|\phi\rangle$
between the ancillas and the meter. If there is no post-selection
then the quantum Fisher information is found to be 
\begin{equation}
I(g)=4\left[\langle\hat{A}^{2}\rangle_{|\Psi_{i}\rangle}\langle\hat{F}^{2}\rangle_{|\phi\rangle}-\left(\langle\hat{A}\rangle_{|\Psi_{i}\rangle}\langle\hat{F}\rangle_{|\phi\rangle}\right)^{2}\right].\label{eq:maxinf}
\end{equation}

Now suppose we projectively measure the ancillas in order to make
a post-selection. This measurement will produce $d^{n}$ independent
outcomes corresponding to some orthonormal basis $\{|\Psi_{f}^{(k)}\rangle\}_{k=1}^{d^{n}}$.
In the linear response regime with $g\ll1$, each of these outcomes
prepares a particular meter state 
\begin{align}
|\phi'_{k}\rangle & \propto\langle\Psi_{f}^{(k)}|\exp(-ig\hat{H})|\Psi_{i}\rangle|\phi\rangle\approx(\hat{1}-igA_{w}^{(k)}\hat{F})|\phi\rangle\label{eq:stateps}
\end{align}
with probability $P_{s}^{(k)}\approx|\langle\Psi_{f}^{(k)}|\Psi_{i}\rangle|^{2}$
that is governed by a different weak value
\begin{equation}
A_{w}^{(k)}=\frac{\langle\Psi_{f}^{(k)}|\hat{A}|\Psi_{i}\rangle}{\langle\Psi_{f}^{(k)}|\Psi_{i}\rangle}.
\end{equation}

We can then compute the remaining Fisher information contained in
each of the post-selected states $\sqrt{P_{s}^{(k)}}|\phi'_{k}\rangle$
using \eqref{eq:2}, which produces 
\begin{align}
I^{(k)}(g) & \approx4\, P_{s}^{(k)}|A_{w}^{(k)}|^{2}\,\left[\text{Var}(\hat{F})_{|\phi\rangle}-\langle\hat{F}^{2}\rangle_{|\phi\rangle}\left(2g\text{Im}A_{w}^{(k)}\langle\hat{F}\rangle_{|\phi\rangle}+|gA_{w}^{(k)}|^{2}\langle\hat{F}^{2}\rangle_{|\phi\rangle}\right)\right].\label{eq:kinf}
\end{align}
Importantly, if we add the information from all $d^{n}$ post-selections
we obtain 
\begin{align}
\sum_{k=1}^{d^{n}}I^{(k)}(g) & \approx4\,\langle\hat{A}^{2}\rangle_{|\Psi_{i}\rangle}\,\text{Var}(\hat{F})_{|\phi\rangle}-O(g).
\end{align}
With the condition $\langle\hat{F}\rangle_{|\phi\rangle}=0$, this
saturates the maximum in \eqref{eq:maxinf} up to small corrections,
which indicates that the ancilla measurement does not lose information
by itself. One can always examine all $d^{n}$ ancilla outcomes to
obtain the maximum information, as pointed out in \cite{Ferrie2013}.

Now let us focus on a particular post-selection $k=1$, using an unbiased
meter that satisfies $\langle\hat{F}\rangle_{|\phi\rangle}=0$, as
assumed in the main text. This produces the simplification 
\begin{align}
I^{(1)}(g) & \approx4\, P_{s}^{(1)}|A_{w}^{(1)}|^{2}\,\left[1-|gA_{w}^{(1)}|^{2}\text{Var}(\hf)\right].\label{eq:kinfsimple}
\end{align}
Now recall Eq.~(15) of the main text, where we showed that if we
fix $P_{s}^{(1)}\ll1$ and picked a post-selection state that maximizes
$A_{w}^{(1)}$ then we found
\begin{equation}
\max|A_{w}^{(1)}|^{2}\approx\frac{1-P_{s}^{(1)}}{P_{s}^{(1)}}\text{Var}(\hat{A})_{|\Psi_{i}\rangle}\approx\frac{\text{Var}(\hat{A})_{|\Psi_{i}\rangle}}{P_{s}^{(1)}}.
\end{equation}
For this strategically chosen post-selection with small $P_{s}^{(1)}$
and maximized $A_{w}^{(1)}$, it then follows that 
\begin{align}
I^{(1)}(g) & \approx4\,\text{Var}(\hat{A})_{|\Psi_{i}\rangle}\,\left[1-|gA_{w}^{(1)}|^{2}\text{Var}(\hf)\right]=I(g)\;\left[\frac{\text{Var}(\hat{A})_{|\Psi_{i}\rangle}}{\langle\hat{A}^{2}\rangle_{|\Psi_{i}\rangle}}\right]\,\left[1-|gA_{w}^{(1)}|^{2}\text{Var}(\hf)\right],\label{eq:kinfopt}
\end{align}
which is Eq.~(16) in the main text. That is, nearly \emph{all} the
Fisher information can be concentrated into a single (but rarely post-selected)
meter state (see also \cite{Jordan2013}). The remaining information
is distributed among the $(d^{n}-1)$ remaining states, and could
be retrieved in principle. The special post-selected meter state suffers
an overall reduction factor of $\eta=\text{Var}(\hat{A})/\langle\hat{A}^{2}\rangle$,
as well as a small loss $|gA_{w}^{(1)}|^{2}\text{Var}(\hf)$. However,
most weak value amplification experiments operate in the linear response
regime $g|A_{w}^{(1)}|\text{Var}(\hf)^{\frac{1}{2}}\ll1$ where this
remaining loss is negligible. Moreover, the overall reduction factor
$\eta$ can even be set to unity by choosing ancilla observables that
satisfy $\langle\hat{A}\rangle_{|\Psi_{i}\rangle}=0$.

As carefully discussed in \cite{Ferrie2013}, one cannot actually
reach the optimal bound of \eqref{eq:maxinf} when making a post-selection.
However, \eqref{eq:kinfopt} shows that one can get remarkably close
by carefully choosing which post-selection to make. It is quite surprising
that one can even approximately saturate \eqref{eq:maxinf} while
discarding the $(d^{n}-1)$ much more probable outcomes. Rare post-selections
can often be advantageous for independent reasons (e.g., to attenuate
an optical beam down to a manageable post-selected beam power), so
this property of weak value amplification makes it an attractive technique
for estimating an extremely small parameter $g$ that permits the
linear response conditions \cite{Jordan2013}.

\subsection{Examples}

To see how this works in more detail, let us examine the ancilla qubit
post-selection examples used in the main text, where $g=\varphi/2$.
For completeness, we will work through two examples. First, we consider
a sub-optimal ancilla observable $\hat{a}=|1\rangle\langle1|$. Second,
we consider an optimal ancilla observable $\hat{a}=\hat{\sigma}_{z}$
to emphasize the practical difference.

\subsubsection{Ancilla Projectors}

A suboptimal choice of ancilla observable is the projector $\hat{a}=|1\rangle\langle1|$
used in controlled qubit operations. From the optimal initial state
given by Eq.~(10) in the main text, we have $\langle\hat{A}^{2}\rangle=n^{2}/2$
and $\langle\hat{A}\rangle=n/2$. Therefore, the maximum quantum Fisher
information from \eqref{eq:maxinf} that we can expect for estimating
$\varphi$ is 
\begin{equation}
I(\varphi)=\frac{n^{2}}{2},\label{eq:4}
\end{equation}
where the factor $1/2$ in $g=\varphi/2$ has been taken into account,
and the corresponding quantum Cram\'{e}r-Rao bound is $\sqrt{2}/n$.
This is the best (Heisenberg) scaling of the estimation precision
that can be obtained by using $n$ entangled ancillas with the given
initial states.

Now, let us consider what happens when we make the optimal preparation
and post-selections for weak value amplification. We expect from \eqref{eq:kinfopt}
that the maximum information which can be attained through post-selection
will be reduced by a factor of
\begin{equation}
\eta=\frac{\text{Var}(\hat{A})_{|\Psi_{i}\rangle}}{\langle\hat{A}^{2}\rangle_{|\Psi_{i}\rangle}}=\frac{1}{2}.
\end{equation}
It is in this sense that the choice of $\hat{a}$ as a projector is
suboptimal. We will see in the next section what happens with the
optimal choice of $\hat{\sigma}_{z}$.

In the first case considered in the main text (i.e., increasing the
post-selection probability with the weak value $A_{w}$ fixed), the
optimal post-selected state is 
\begin{equation}
|\Psi_{f}\rangle\propto(A_{w}^{*})|1\rangle^{\otimes n}+(n-A_{w}^{*})|0\rangle^{\otimes n}.\label{eq:post}
\end{equation}
Computing the post-selected meter state then produces 
\begin{equation}
|\phi'\rangle_{1}=\frac{\left[n-A_{w}[1-\cos(n\varphi/2)]\hat{1}-iA_{w}\sin(n\varphi/2)\hat{\sigma}_{z}\right]|\phi\rangle}{\left(n^{2}+2[|A_{w}|^{2}-n\text{Re}A_{w}][1-\cos(n\varphi/2)]\right)^{1/2}}\approx\left(\hat{1}-iA_{w}\frac{\varphi}{2}\hat{\sigma}_{z}\right)|\phi\rangle,
\end{equation}
where we have used $\langle\phi|\hat{\sigma}_{z}|\phi\rangle=0$,
and then have made the small parameter approximation $n\varphi\ll1$.
This recovers the expected linear response result in \eqref{eq:stateps}.
This state is post-selected with probability 
\begin{equation}
p_{1}=\frac{1}{2}-\cos(n\varphi/2)\frac{|A_{w}|^{2}-n\text{Re}A_{w}}{n^{2}+2[|A_{w}|^{2}-n\text{Re}A_{w}]}\approx\frac{n^{2}}{2n^{2}+4[|A_{w}|^{2}-n\text{Re}A_{w}]}\approx\frac{n^{2}}{4}|A_{w}|^{-2},
\end{equation}
where we have made the small parameter approximation $n\varphi\ll1$,
and then the large weak value assumption $n\ll|A_{w}|$.

Now computing the quantum Fisher information (\ref{eq:2}) with the
post-selected meter state $\sqrt{p_{1}}\,|\phi'\rangle_{1}$ yields
the simple expression 
\begin{equation}
I_{1}(\varphi)\approx\frac{n^{2}}{4}\left(1-\left|\frac{\varphi A_{w}}{2}\right|^{2}\right)\leq\frac{n^{2}}{4},\label{eq:fisher1}
\end{equation}
in agreement with \eqref{eq:kinfopt}. The maximum achieves the best
possible scaling of $n^{2}$ as in \eqref{eq:4}. Moreover, for the
most frequently used linear response regime with $|A_{w}|\varphi\ll1$,
we achieve the expected maximum information of $\eta I(\varphi)=n^{2}/4$.

For the second case (i.e., increasing the weak value $A_{w}$ with
the post-selection probability fixed), we can obtain the results simply
by rescaling $A_{w}\to\sqrt{n}A_{w}$ to produce $p_{2}\propto n$,
as shown in the main text. This produces, 
\begin{equation}
|\phi'\rangle_{2}\approx\left(\hat{1}-i\sqrt{n}A_{w}\frac{\varphi}{2}\hat{\sigma}_{z}\right)|\phi\rangle,
\end{equation}
and 
\begin{equation}
p_{2}\approx\frac{n^{2}}{4}|\sqrt{n}A_{w}|^{-2}=\frac{n}{4}|A_{w}|^{-2},
\end{equation}
and yields the Fisher information 
\begin{equation}
I_{2}(\varphi)\approx\frac{n^{2}}{4}\left(1-n\left|\frac{\varphi A_{w}}{2}\right|^{2}\right)\leq\frac{n^{2}}{4}.\label{eq:fisher2}
\end{equation}
The increase of the amplification factor $|A_{w}|$ correspondingly
decreases the remaining Fisher information, as expected from \eqref{eq:fisher1}.
However, since $n\varphi\ll1$ and $\varphi|A_{w}|\ll1$ in the linear
response regime, this decrease is still small.

Alternatively, this second case can be computed explicitly as follows.
For a fixed post-selection probability $p$, the post-selected state
must be $|\Psi_{f}\rangle=\sqrt{p}|\Psi_{i}\rangle+\sqrt{1-p}|\Psi_{i}^{\perp}\rangle,$
where the optimal $|\Psi_{i}^{\perp}\rangle$ is parallel to the component
of $\ha|\Psi_{i}\rangle$ in the complementary subspace orthogonal
to $|\Psi_{i}\rangle$. Computing this yields 
\begin{equation}
\begin{aligned}|\Psi_{f}\rangle & =\sqrt{p}|\Psi_{i}\rangle+\sqrt{1-p}\frac{\hat{A}|\Psi_{i}\rangle-|\Psi_{i}\rangle\langle\Psi_{i}|\hat{A}|\Psi_{i}\rangle}{\sqrt{\text{Var}(\hat{A})_{|\Psi_{i}\rangle}}}\\
 & =\left(\sqrt{\frac{p}{2}}-\sqrt{\frac{1-p}{2}}\right)|0\rangle^{\otimes n}+\left(\sqrt{\frac{p}{2}}+\sqrt{\frac{1-p}{2}}\right)|1\rangle^{\otimes n}.
\end{aligned}
\label{eq:fixp}
\end{equation}
Thus, computing the post-selected meter state yields 
\begin{equation}
|\phi'\rangle_{2}\propto\left(\left(\sqrt{\frac{p}{2}}-\sqrt{\frac{1-p}{2}}\right)\hat{1}+\left(\sqrt{\frac{p}{2}}+\sqrt{\frac{1-p}{2}}\right)\e^{-in\varphi\hat{\sigma}_{z}/2}\right)|\phi\rangle\approx\left(\hat{1}-i|A_{w}|\frac{\varphi}{2}\hat{\sigma}_{z}\right)|\phi\rangle,
\end{equation}
where we have defined the effective weak value factor 
\begin{equation}
|A_{w}|=\frac{n}{2}\left(1+\sqrt{\frac{1-p}{p}}\right)\approx\frac{n}{2}p^{-1/2},\label{eq:effaw}
\end{equation}
and have used the linear response approximations $n\varphi\ll1$ and
$\varphi|A_{w}|\ll1$, as well as the small probability assumption
$p\ll1$. Computing the quantum Fisher information from (\ref{eq:2})
with the state $\sqrt{p}\,|\phi'\rangle_{2}$ then produces 
\begin{equation}
I_{2}(\varphi)\approx p|A_{w}|^{2}\left(1-\left[\frac{\varphi|A_{w}|}{2}\right]^{2}\right)=\frac{n^{2}}{4}\left(1-\left[\frac{n\varphi}{4\sqrt{p}}\right]^{2}\right)\leq\frac{n^{2}}{4}
\end{equation}
using the definition \eqref{eq:effaw}. This result precisely matches
the form of \eqref{eq:kinfsimple}. It is now clear that for quadratic
scaling $p=n^{2}p_{0}$ we recover \eqref{eq:fisher1} with the effective
reference weak value $|A_{w}|=1/(2\sqrt{p_{0}})$, while for linear
scaling $p=np_{0}$ we recover \eqref{eq:fisher2}.

\subsubsection{Ancilla Z-operators}

For contrast, an optimal choice of ancilla observable is $\hat{a}=\hat{\sigma}_{z}$,
as used in the main text. From the optimal initial state given by
Eq.~(10) in the main text, we have $\langle\hat{A}^{2}\rangle=n^{2}$
and $\langle\hat{A}\rangle=0$. Therefore, the maximum quantum Fisher
information from \eqref{eq:maxinf} that we can expect for estimating
$\varphi$ is 
\begin{equation}
I(\varphi)=n^{2},\label{eq:4b}
\end{equation}
which is a factor of 2 larger than \eqref{eq:4}. The corresponding
quantum Cram\'{e}r-Rao bound is $1/n$. From \eqref{eq:kinfopt},
we expect that the reduction factor is
\begin{equation}
\eta=\frac{\text{Var}(\hat{A})_{|\Psi_{i}\rangle}}{\langle\hat{A}^{2}\rangle_{|\Psi_{i}\rangle}}=1.
\end{equation}
Thus, it is possible to saturate the optimal bound with this choice
of $\hat{a}$.

In the first case considered in the main text (i.e., increasing the
post-selection probability with the weak value $A_{w}$ fixed), the
optimal post-selected state is 
\begin{equation}
|\Psi_{f}\rangle\propto(n+A_{w}^{*})|1\rangle^{\otimes n}+(n-A_{w}^{*})|0\rangle^{\otimes n}.\label{eq:postz}
\end{equation}
Computing the post-selected meter state then produces 
\begin{equation}
|\phi'\rangle_{1}=\frac{\left[n\cos(n\varphi/2)\hat{1}-iA_{w}\sin(n\varphi/2)\hat{\sigma}_{z}\right]|\phi\rangle}{\left(n^{2}\cos^{2}(n\varphi/2)+|A_{w}|^{2}\sin^{2}(n\varphi/2)\right)^{1/2}}\approx\left(\hat{1}-iA_{w}\frac{\varphi}{2}\hat{\sigma}_{z}\right)|\phi\rangle,
\end{equation}
where we have used $\langle\phi|\hat{\sigma}_{z}|\phi\rangle=0$,
and then have made the small parameter approximation $n\varphi\ll1$.
This again recovers the expected linear response result in \eqref{eq:stateps}.
This state is post-selected with probability 
\begin{equation}
p_{1}=\frac{n^{2}\cos^{2}(n\varphi/2)+|A_{w}|^{2}\sin^{2}(n\varphi/2)}{n^{2}+[A_{w}|^{2}}\approx\frac{n^{2}}{n^{2}+|A_{w}|^{2}}\approx n^{2}|A_{w}|^{-2},
\end{equation}
where we have made the small parameter approximation $n\varphi\ll1$,
and then the large weak value assumption $n\ll|A_{w}|$.

Now computing the quantum Fisher information (\ref{eq:2}) with the
post-selected meter state $\sqrt{p_{1}}\,|\phi'\rangle_{1}$ yields
the simple expression 
\begin{equation}
I_{1}(\varphi)\approx n^{2}\left(1-\left|\frac{\varphi A_{w}}{2}\right|^{2}\right)\leq n^{2},\label{eq:fisher1z}
\end{equation}
in agreement with \eqref{eq:kinfopt}. The maximum saturates the upper
bound of $n^{2}$ in \eqref{eq:4b}, as expected.

For the second case (i.e., increasing the weak value $A_{w}$ with
the post-selection probability fixed), we can again obtain the results
simply by rescaling $A_{w}\to\sqrt{n}A_{w}$ to produce 
\begin{align}
|\phi'\rangle_{2} & \approx\left(\hat{1}-i\sqrt{n}A_{w}\frac{\varphi}{2}\hat{\sigma}_{z}\right)|\phi\rangle,\\
p_{2} & \approx n^{2}|\sqrt{n}A_{w}|^{-2}=n|A_{w}|^{-2},
\end{align}
and the Fisher information 
\begin{equation}
I_{2}(\varphi)\approx n^{2}\left(1-n\left|\frac{\varphi A_{w}}{2}\right|^{2}\right)\leq n^{2}.\label{eq:fisher2z}
\end{equation}

Alternatively, computing the optimal post-selection state for a fixed
post-selection probability $p$ yields the same state as \eqref{eq:fixp}.
Hence, computing the post-selected meter state yields 
\begin{equation}
|\phi'\rangle_{2}\propto\left(\left(\sqrt{\frac{p}{2}}-\sqrt{\frac{1-p}{2}}\right)e^{in\varphi\hat{\sigma}_{z}/2}+\left(\sqrt{\frac{p}{2}}+\sqrt{\frac{1-p}{2}}\right)\e^{-in\varphi\hat{\sigma}_{z}/2}\right)|\phi\rangle\approx\left(\hat{1}-i|A_{w}|\frac{\varphi}{2}\hat{\sigma}_{z}\right)|\phi\rangle,
\end{equation}
where we have defined the effective weak value factor 
\begin{equation}
|A_{w}|=n\sqrt{\frac{1-p}{p}}\approx np^{-1/2},\label{eq:effaw2}
\end{equation}
in contrast to \eqref{eq:effaw}. Computing the quantum Fisher information
from (\ref{eq:2}) with the state $\sqrt{p}\,|\phi'\rangle_{2}$ then
produces 
\begin{equation}
I_{2}(\varphi)\approx p|A_{w}|^{2}\left(1-\left[\frac{\varphi|A_{w}|}{2}\right]^{2}\right)=n^{2}\left(1-\left[\frac{n\varphi}{\sqrt{p}}\right]^{2}\right)\leq n^{2},
\end{equation}
using the definition \eqref{eq:effaw2}. As before, this result precisely
matches the form of \eqref{eq:kinfsimple}. It is now clear that for
quadratic scaling $p=n^{2}p_{0}$ we recover \eqref{eq:fisher1z}
with the effective reference weak value $|A_{w}|=1/\sqrt{p_{0}}$,
while for linear scaling $p=np_{0}$ we recover \eqref{eq:fisher2z}.
Therefore, in both post-selected qubit examples considered in the
main text we can nearly saturate the expected maximum of $I(\varphi)=n^{2}$
when the linear response conditions $n\varphi\ll1$, $\varphi|A_{w}|\ll1$,
and the large weak value condition $n\ll|A_{w}|$ are met, despite
the loss of data incurred by the post-selection.

\end{widetext}

\begin{thebibliography}{10}
\bibitem{footnote}Note that some references have also considered
higher precision scalings such as $n^{-k}$ that arise when there
are $k$-body interactions between the $n$ ancillas \cite{quantum metrology 3}.
\end{thebibliography}
\end{document}